\documentclass[prl,letterpaper,twocolumn,showpacs,superscriptaddress,amsmath,amsfonts,amssymb,preprintnumbers]{revtex4}
\pdfoutput=1
\usepackage[T1]{fontenc}\usepackage[latin1]{inputenc}
\usepackage{dcolumn,graphicx,color,booktabs,microtype}
\usepackage[charter]{mathdesign}
\renewcommand{\figurename}{Fig.}
\renewcommand{\tablename}{Table}
\makeatletter\renewcommand{\fnum@figure}[1]{\figurename~\thefigure.}\makeatother
\makeatletter\renewcommand{\fnum@table}[1]{\tablename~\thetable.}\makeatother

\usepackage[colorlinks,plainpages=false,linkcolor=black,urlcolor=blue,citecolor=black,pdfpagemode=UseNone,pdfstartview=FitBH]{hyperref}

\begin{document}

\title{Evidence for Fermi surface reconstruction in the static stripe phase of La$_{1.8-x}$Eu$_{0.2}$Sr$_x$CuO$_{4}$,~$x=1/8$}

\author{V.~B.~Zabolotnyy}
\affiliation{Institute for Solid State Research, IFW-Dresden,
P.O.Box 270116, D-01171 Dresden, Germany}

\author{A.~A.~Kordyuk}
\affiliation{Institute for Solid State Research, IFW-Dresden,
P.O.Box 270116, D-01171 Dresden, Germany} \affiliation{Institute
of Metal Physics of National Academy of Sciences of Ukraine, 03142
Kyiv, Ukraine}
\author{D.~S.~Inosov}
\author{D.~V.~Evtushinsky}
\author{R.~Schuster}
\author{B.~Büchner}
\author{N.~Wizent}
\author{G.~Behr}
\affiliation{Institute for Solid State Research, IFW-Dresden, P.O.Box 270116, D-01171 Dresden,
Germany}
\author{Sunseng Pyon}
\author{H. Takagi}
\address{Engineering Research Institute, University of Tokyo, Yayoi, Bunkyo-ku, Tokyo 113, Japan}

\author{R.~Follath}
\address{BESSY GmbH, Albert-Einstein-Strasse 15, 12489 Berlin, Germany}

\author{S.~V.~Borisenko}
\affiliation{Institute for Solid State Research, IFW-Dresden, P.O.Box 270116, D-01171 Dresden,
Germany}


\begin{abstract}
We present a photoemission study of La$_{0.8-x}$Eu$_{0.2}$Sr$_x$CuO$_{4}$  with doping level
$x$=1/8, where the charge carriers are expected to order forming  static stripes. Though the local
probes in direct space seem to be consistent with this idea, there has been little evidence found
for such ordering in quasiparticle dispersions. We show that the Fermi surface topology of the 1/8
compound develops notable deviations from that observed for La$_{2-x}$Sr$_x$CuO$_{4}$  in a way
consistent with the FS reconstruction expected for the scattering on the antiphase stripe order.
\end{abstract}

\maketitle

Since the discovery of charge- and spin-ordering in high-$T_\textup{c}$ cuprates, the phenomenon
has attracted much attention, both from the theoretical and experimental point of view
\cite{Kivelson, Castro}. Moreover, there appeared a number of theoretical approaches  considering
the charge and spin segregation as having strong, if not decisive, impact on the onset of
superconductivity in high-$T_\textup{c}$ superconductors \cite{Vojta,Martin}. The ordering effects
were found  to depend crucially on the charge doping level, being the most pronounced for the
pseudogap regime in the vicinity of doping level $x=1/8$, where the doped holes are expected to
form so-called stripes with the antiferromagnetically ordered spins. Generally, the stripe order is
supposed to fluctuate, though for particular superconductors like La$_{2-x-y}$M$_x$Sr$_y$CuO$_4$ (M
= Nd or La) the inhomogeneities were shown to be practically static \cite{Tranquada, Klauss,
Kojima, Hunt, Teitelbaum, Kataev, Suh, Simovic}, making those compounds most prevalent in
experimental research. The spin response of the stripe phase has been studied in inelastic neutron
scattering experiments \cite{Tranquada, Tranquada2} supporting the idea of spin ordering.
Similarly, local probes, such as scanning tunnelling microscopy, have clearly demonstrated charge
modulation on the surface of high temperature superconductors \cite{Hoffman, Vershinin}, changing
the hypothesis of spin and charge modulation into a well established fact. On general grounds any
charge/spin ordering must act as an additional scattering potential, resulting in a reconstruction
of the initial Fermi surface (FS). Indirect evidence for such modifications comes from Hall
coefficient and de Haas-van Alphen measurements \cite{Leboeuf, Adachi, Nakamura, Suchitra},
suggesting a formation of new orbits when the stripe order sets in. Nonetheless the experiments
that would explicitly expose the effect of charge stripe order on the free charge carriers, namely
modifications to the electronic band dispersion and topology of the FS, are not numerous and
suggest radically different distribution of quasiparticle spectral weight over the Brillouin zone
\cite{Zhou, Valla}.

Here we present experimental data on the electronic band structure of
La$_{1.675}$Eu$_{0.2}$Sr$_{0.125}$CuO$_4$ obtained using angle resolved  photoelectron spectroscopy
(ARPES) and compare the topology of the experimentally observed FS to that of pure
La$_{2-x}$Sr$_x$CuO$_{4}$ (LSCO) samples as well as to predictions obtained within a simple model
where electrons scatter on effective potential induced by the stripe and charge order. We show that
the measured distribution of photoelectron intensity is consistent with the FS reconstruction
expected for the antiphase stripe order \cite{Zaanen, Millis} and give quantitative estimates for
the strength of the scattering potential in the spin and charge channels.
\begin{figure}
\begin{center}
\includegraphics[width=\columnwidth]{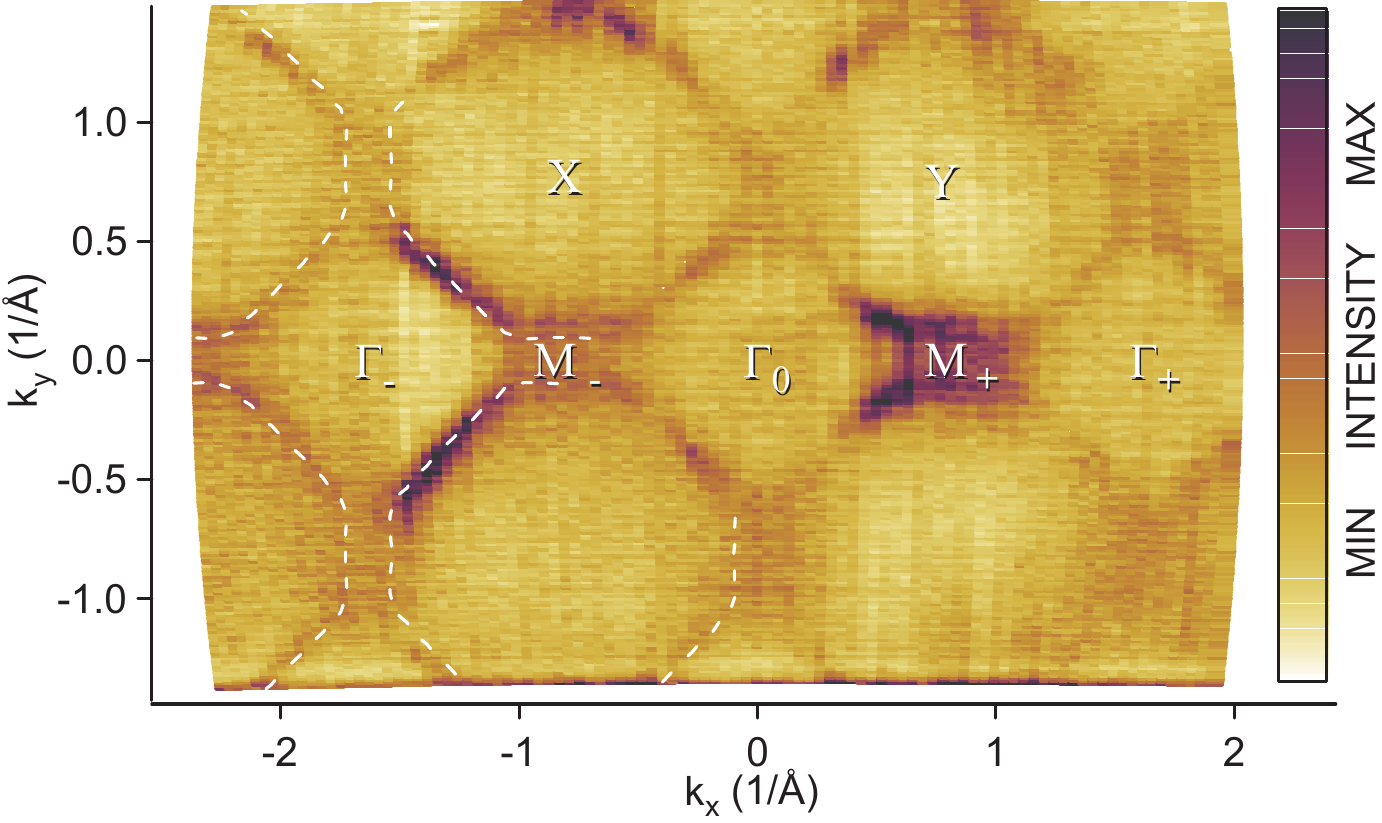}\vspace{-1em}
\caption{Experimental FS map of La$_{1.675}$Eu$_{0.2}$Sr$_{.125}$CuO$_4$, $T=25 $ K. No
symmetrization was applied, the map contains a set of independent $\mathbf{k}$ points covering
several Mahan cones\,\cite{Mahan}. The map was measured with light polarization perpendicular to
the analyzer entrance slit, and normalized to the total intensity. Further details on experimental
geometry and data processing  can be found elsewhere \cite{Zabolotnyy,Inosov,
Borisenko2}.\vspace{-1.5em}}
\end{center}
\end{figure}

\begin{figure*}[t]
\begin{center}\vspace{-0.3em}
\includegraphics[width=0.85\textwidth]{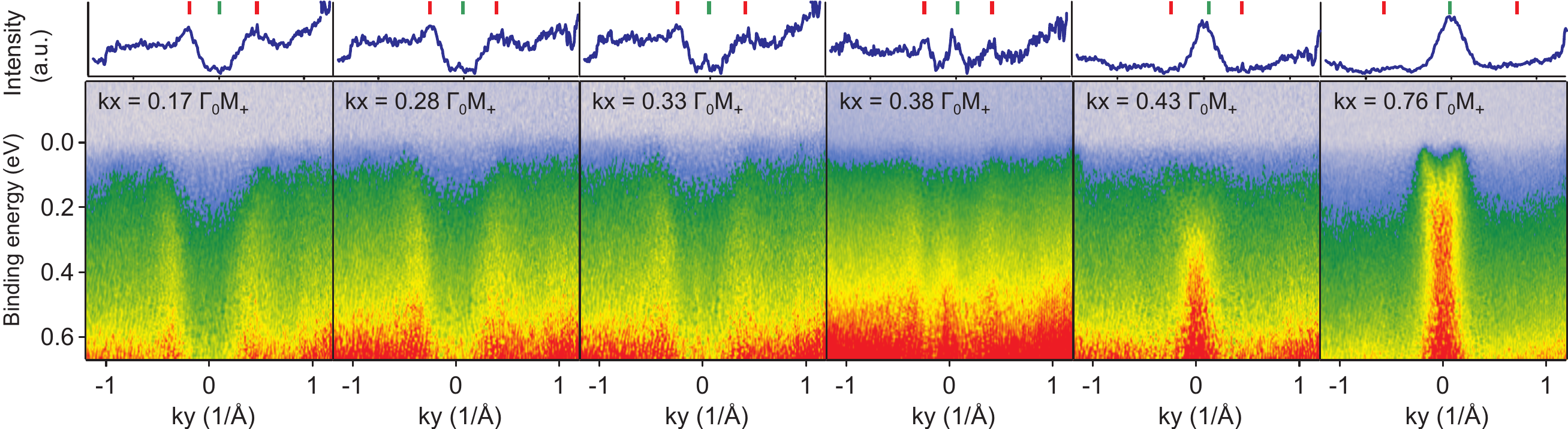}\vspace{-0.5em}\\
\caption{Evolution of the spectral weight through the Brilloin zone.  The curves on top of each
energy-momentum image represents an MDC integrated in the energy window 0.3\,--\,0.4
eV.\vspace{-1.5em}}
\end{center}
\end{figure*}

The experimental data were collected using  1$^3$ station at the BESSY synchrotron facility with
energy and momentum resolution of 12\,meV and 0.05\AA $^{-1}$ respectively. The high quality single
crystals of La$_{1.675}$Eu$_{0.2}$Sr$_{0.125}$CuO$_4$ with suppressed superconductivity were
mounted on the cryomanipulator and cleaved \textit{in situ} in ultrahigh vacuum. All the data
presented in the manuscript were collected at low temperature $T=25$\,K.

We start the discussion of the experimental data with the FS map plotted in Fig.\,1, which
represents the photoelectron intensity integrated over a small energy window $E =
E_\textup{F}\pm15$\,meV.  While for pure LSCO with the  doping level $0.05 \lesssim x \lesssim 0.17
$ the FS consists of the rounded  contours centered at the X/Y points, for the Eu-doped sample the
form of the FS contours is qualitatively different. There are extended and practically straight FS
segments passing through the nodal points comprising
 45$^\circ$ angle with the primary axes. At the antinodal point the apparent FS contour changes direction, forming
segments parallel to the primary crystallographic axes, so that the whole FS rather reminds an
octagon as shown by dotted guide lines in Fig.\,1.

Assuming that the observed contours represent a true connected FS we tried to fit it with a
standard tight-binding (TB) formula\cite{Yoshida}. It was practically impossible to find a set of
TB parameters that would provide a reasonable fit both at the Fermi level (FL) as well as at
higher binding energies.  This was the first indirect indication that the assumed topology of the
FS is not a true one, in a sense that the apparent FS may consist of several disjoint pockets,
which in view of disorder introduced by Eu doping or/and short correlation length of the stripe
potential are hard to detect.
 In some sense the situation may be similar to the electron doped cuprates, whose FS  undergoes  reconstruction with
 decreasing the charge doping,   though in the photoemission data the  reconstructed  FS consisting of the hole  and electron pockets still resembles the
 unreconstructed one \cite{PCCO}.

  Conjecture that the  embraced by the  guide line area corresponds to the hole doped region would also results in  the exaggerated doping
$x=0.19\pm0.02$ and thus would be at variance with a reasonable agreement between the nominal
doping level of the LSCO system and the one estimated by the FS area \cite{Yoshida}. Other, and
probably  the most straightforward, evidence for the discontinuity of the assumed FS contour comes
from the analysis of the photoemission intensity over the whole measured range of binding energies.
In Fig.\,2 we plot a series of energy-momentum cuts, representing photoelectron intensity for
several fixed $k_x$ values spanning from $\Gamma_0$ to $\textup{M}_1$ point as a function of $k_y$
and binding energy.  In the first column one can clearly see two bands crossing the FL at
$k_y\backsimeq\pm 0.3 $\AA$^{-1}$. Were the assumed octagonal FS contours really continuous, one
should see these two bands and corresponding FL crossings gradually moving closer to each other as
$k_x$ approaches the M point. Indeed the aforementioned  FL crossings are getting slightly closer
as it follows from the second column of Fig.\,2, but at $k_y = 0$ there appears another band
 that gains the intensity and finally results in a well defined FL crossing  for the  $k_x
\gtrsim 0.6 \Gamma_0\textup{M}_{+}$, i.e. the FS segments at $k_x \lesssim 0.6
\Gamma_0\textup{M}_{+}$ and $k_x \gtrsim 0.6 \Gamma_0\textup{M}_{+}$ must belong to a separate FS
sheet and the seeming continuity must be only due to the large momentum width of the features and
specifics of the photoemission that make the FS segments that practically coincide with  the
``parent'' FS the most intense and dominating over the replicas in the ARPES signal.

Along with the FS breaking into several sheets one would also expect  characteristic for this case
backfolding effects in the band dispersions.  Nevertheless this kind of reasoning must be exploited
cautiously as the coherence length of the folding potential may result in notable deviation from
the simple picture. It is well known that the FS of some electron doped cuprates is reconstructed,
splitting into the hole and electron pockets \cite{PCCO}. The reason for this reconstruction is
likely  to be the short range antiferromagnetic correlations that appear for the electron doping
$x\lesssim0.14$. In Fig. 3\,(a) we show a typical ``backfolded''  band for
Pr$_{1.85}$Ce$_{0.15}$CuO$_4$. As can be seen, the major signature of reconstruction can be
described as the band having been ``chopped off'' above the line A-A (Fig. 3a).  Analogous effects
can also be found in  the energy-momentum intensity distribution for the Eu-LSCO sample [see Fig.
3(b)] thus providing another indication for the FS  reconstruction.
\begin{figure}
\begin{center}
\includegraphics[width=0.7\columnwidth]{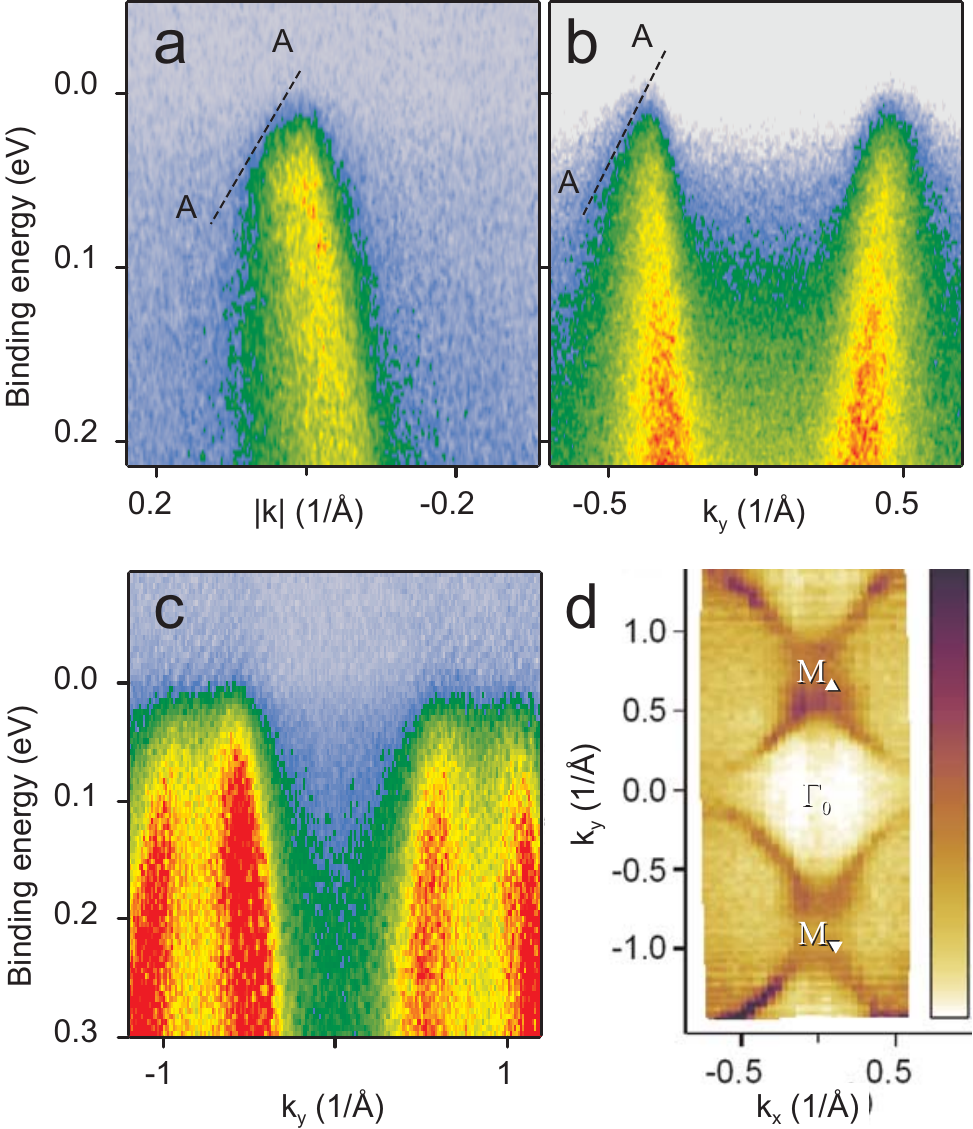}\vspace{-0.5em}\\
\caption{Effect of the band folding in the spectral function of  Pr$_{1.85}$Ce$_{0.15}$CuO$_4$ (a)
and  Eu-LSCO (b). High symmetry cut $\Gamma_\blacktriangle -\Gamma_0 -
\Gamma_\blacktriangledown$\,(c) and FS that are supporting a possible FL band crossing along the
$\Gamma_\blacktriangle -\Gamma_0 - \Gamma_\blacktriangledown$ direction. Unlike all the other, the
data in panels (c)--(d) were obtained  with the inplane component of the light polarization vector
parallel to the $k_y$ axis.\vspace{-3.0em} }
\end{center}
\end{figure}
Besides the discussed signatures  of the FS reconstruction, there is also a seeming
one-dimensionality in the  experimental data. In particular, in Fig.\,1 the FS segments parallel to
the $k_x$ axis are well pronounced, while their counterparts, which are supposed to be parallel to
the $k_y$ axis, are rather diffuse. To check if  this might be an evidence for a generic anisotropy
in the electronic structure or a mere effect of photoemission matrix elements, we have made similar
measurements with the light polarization rotated by 90$^\circ$. The corresponding FS map and  the
hight symmetry cut $\Gamma_\blacktriangle -\Gamma_0 - \Gamma_\blacktriangledown$ are shown in Figs.
3(c)\,--\,(d). As can be seen in Fig.\,3c the photoemission intensity still remains suppressed in
the vicinity of $\Gamma_{\blacktriangle/\blacktriangledown}$ points, which is rather unusual if
attributed to unfavorable matrix elements. Moreover, the observed intensity distribution in the
$\Gamma_\blacktriangle -\Gamma_0 - \Gamma_\blacktriangledown$ cut is more consistent with band
crossing the FL, rather than with a saddle point expected for stripe-free LSCO system.

 One of the first easy-to-grasp consideration of the FS reconstructions due to the
spin and charge modulation has been given in Ref.\,\onlinecite{Millis}, but focusing mainly on the
Hall effect the authors restricted the discussion to the formal FS topology. When aiming at
comparison to the photoemission experiment such a description needs to be extended, since these are
not the bands that are seen directly in the photoemission experiment but the spectral function
modified by the photoemission matrix elements $\Delta_{f,i}$. Even when the matrix elements can be
neglected in the unreconstructed case, the intensity variations upon the reconstruction are
generally tremendous \cite{Borisenko}. Typically one notes a significantly lower intensity of the
newly sprang-up replicas as compared to the original unreconstructed bands. To understand this
important and ubiquitous intensity disparity \cite{Borisenko, Brouet} one may turn to a sudden
approximation as the simplest, but still sufficient for this purpose, approach. In this case the
 intensity of photoelectrons  detected at some final state $|f \rangle$ can be written as \cite{Hedin}:
\begin{equation}
J\!_f(\omega) \!=\! f(\omega)\!\sum_{i}|\Delta_{f,i}|^2\!A_i(\omega)\!= \!\sum_{i}\!|\Delta_{f,i}|^2
A^{<}_i(\omega).\vspace{-0.5em}
\end{equation}
Here the summation runs over the set of one-electron states $|i\rangle$ forming the basis in which
the spectral function $A_i(\omega)$ is given. For our particular purpose, when considering a 2D
cuprate, we reduce the complete basis set to  states forming  the single band crossing the FL, and
naturally enumerate them by the Bloch quasi-momentum $\textbf{k} = (k_x, k_y)$ limited to the
unreconstructed BZ: $k_x\in[-\frac{\pi}{a};+\frac{\pi}{a} )$ ,
$k_y\in[-\frac{\pi}{b};+\frac{\pi}{b} )$. Furthermore, we consider the excitations as  a well
defined quasiparticles and write the Hamiltonian in a diagonal form
\begin{equation}
\hat{H}_0 =\sum_{\mathbf{k}\in \textup{BZ}} \varepsilon_\mathbf{k}\hat{c}_\mathbf{k}^\dag\hat{c}_\mathbf{k}^{\phantom{\dag}},
\end{equation}
with $\varepsilon_\mathbf{k}$ being the renormalized band dispersion. It is easy to check that the
spectral function reduces to a trail of delta functions aligned along the band dispersion
\cite{Mahan_Book}, and that the matrix element $\Delta_{f,i}$ between the final state $|f\rangle$
characterized by the electron momentum $\mathbf{p}$ and initial state $|i\rangle$ given by the
Bloch  wave with quasi-momentum $\mathbf{k}$ ensures periodic replication of the photoemission
picture determined by $A_\mathbf{k}(\omega)$ over different Mahan cones \cite{Mahan} as it can
actually be seen in Fig.\,1.

As was pointed out in Ref. \onlinecite{Millis}, commensurate stripe order assumed  in the model of antiphase stripe
 domains \cite{Zaanen}  induces additional potential due to scattering on spin and charge modulations  $V = V_\textup{s} +
V_\textup{c}$ that can be characterised by two most significant matrix elements
\begin{equation}
\begin{split}
V_\textup{s} = \langle \mathbf{k}|\hat{V}_\textup{s}(\mathbf{r})|\mathbf{k}\pm\mathbf{Q_\textup{s}}\rangle, \textup{with } \mathbf{Q_\textup{s}} = (3\pi/4;\pi ), \textup{ and}\\
V_\textup{c} = \langle \mathbf{k}|\hat{V}_\textup{c}(\mathbf{r})|\mathbf{k}\pm\mathbf{Q_\textup{c}}\rangle, \textup{with } \mathbf{Q_\textup{c}} = (\pi/4; 0 ).
\end{split}
\end{equation}
The choice of this particular model is motivated by the recent inelastic X-ray and neutron
scattering experiments that seem to  be in agreement with the theoretically expected vectors of
charge modulations, not taking into account a small ($\sim$\,3\%) incommensurability \cite{Kim}.

Introducing a quasi-momentum $\mathbf{q}$ limited to the reduced Brilloin zone (RBZ) because of  the enlarged unit cell
in the real space, and zone number $m=0,...,7$, which becomes necessary to describe the states of the original band if
the RBZ is used, the system Hamiltonian for the modulated case can be written as a matrix with respect to the zone index
$m$:\hspace{-0.5em}
\begin{multline}
\hat{H} = \!\!\!\!\!\!\!\! \sum_{\begin{smallmatrix}\mathbf{q}\in \textup{RBZ}\\ m, n =0,...,7 \end{smallmatrix}}
\!\!\!\!\!\!\!\!(\delta_{m,n}\varepsilon_{\mathbf{q}+\mathbf{g}_m} + V_{m,n}) \hat{c
}_{\mathbf{q}+\mathbf{g}_m}^\dag\hat{c}_{\mathbf{q}+\mathbf{g}_n}^{\phantom{\dag}}, \textup{with}
\\V_{m,n}(\mathbf{q})=
       \left(
       \begin{smallmatrix}
       0 & V_\textup{c} & 0 & V_\textup{c} & 0 & V_\textup{s} & V_\textup{s} & 0\\
       V_\textup{c} & 0 & V_\textup{c} & 0 & 0 & 0 & V_\textup{s} & V_\textup{s}\\
       0 & V_\textup{c} & 0 & V_\textup{c} & V_\textup{s} & 0 & 0 & V_\textup{s}\\
       V_\textup{c} & 0 & V_\textup{c} & 0 & V_\textup{s} & V_\textup{s} & 0 & 0\\
       0 & 0 & V_\textup{s} & V_\textup{s} & 0 & V_\textup{c} & 0 & V_\textup{c}\\
       V_\textup{s} & 0 & 0 & V_\textup{s} & V_\textup{c} & 0 & V_\textup{c} & 0\\
       V_\textup{s} & V_\textup{s} & 0 & 0 & 0 & V_\textup{c} & 0 & V_\textup{c}\\
       0 & V_\textup{s} & V_\textup{s} & 0 & V_\textup{c} & 0 & V_\textup{c} & 0\\
       \end{smallmatrix}
       \right),\hspace{2em}
\end{multline}
where $\mathbf{g}_m$ is a set of 8 vectors  determined by the condition $\mathbf{g}_m = k_m \mathbf{Q}_\textup{c} + l_m
\mathbf{Q}_\textup{s}\in \textup{BZ}$ and $k_m, l_m \in \mathbb{Z}$.
 Diagonalizing the matrix $H_{m,n}(\mathbf{q})=
\delta_{m,n}\varepsilon_{\mathbf{q}+\mathbf{g}_m} + V_{m,n} $  results in 8 eigenvalues and
eigenstates for each particular $\mathbf{q}\in \textup{RBZ}$:\vspace{-0.2em}
\begin{equation}
\begin{split}
\hat{H} = \!\!\!\!\!\!\!\! \sum_{\begin{smallmatrix}\mathbf{q}\in \textup{RBZ}\\ i,m, n =0,...,7 \end{smallmatrix}} \!\!\!\!\!\!\!\!D^*_{m, i}(\mathbf{q})\, E_i(\mathbf{q}) \, D_{i, n}(\mathbf{q})
\hat{c}_{\mathbf{q}+\mathbf{g}_m}^\dag\hat{c}_{\mathbf{q}+\mathbf{g}_n}^{\phantom{\dag}}=\\
\sum_{\begin{smallmatrix}\mathbf{q}\in \textup{RBZ}\\ i=0,...,7 \end{smallmatrix}}
\!\!\!\!E_i(\mathbf{q})\hat{a}_{\mathbf{q}, i}^\dag\hat{a}_{\mathbf{q}, i}^{\phantom{\dag}},
\textup{ where } \hat{a}_{\mathbf{q}, i}^{\phantom{\dag}} = \sum_n D_{i,n}(\mathbf{q})  \hat{c}_{\mathbf{q}+\mathbf{g}_n}^{\phantom{\dag}}.
\end{split}
\end{equation}\vspace{-0.5em}

Now, knowing the eigenstates $\hat{a}_{\mathbf{q}, i}^{\dag}|0\rangle$ and eigenenergies
$E_i(\mathbf{q})$  of the reconstructed system we write the spectral function\vspace{-0.3em}
\begin{equation}
\begin{split}
A^{<}_\mathbf{k}(\omega) = \!\!\! \sum_{\begin{smallmatrix}\mathbf{q}\in \textup{RBZ},\\ i=0,...,7 \end{smallmatrix}}\!\!\!\!
\left| \langle 0 | \hat{c}_{\mathbf{k}}^{\phantom{\dag}}|\mathbf{q}, i \rangle \right |^2
\delta(E_i(\mathbf{q}) - \omega) = \\
\!\!\!\!\!\!\!\! \sum_{\begin{smallmatrix}\mathbf{q}\in \textup{RBZ},\\i,m=0,...,7 \end{smallmatrix}} \!\!\!\!\!\!
\left| \langle 0 | \hat{c}_{\mathbf{k}}^{\phantom{\dag}} D^*_{m,i}(\mathbf{q})  \hat{c}_{\mathbf{q}+\mathbf{g}_m}^{\dag} |0 \rangle \right |^2
\delta(E_i(\mathbf{q}) - \omega) = \\
\!\!\!\!\!\!\! \sum_{\begin{smallmatrix}\mathbf{q}\in \textup{RBZ},\\i,m=0,...,7 \end{smallmatrix}} \!\!\!\!\!\!
\left|  D^*_{m,i}(\mathbf{k}- \mathbf{g}_m) \right |^2
\delta(E_i(\mathbf{\mathbf{k}- \mathbf{g}_m}) - \omega).\vspace{-0.2em}
\end{split}
\end{equation}

\begin{figure}
\begin{center}
\includegraphics[width=0.8\columnwidth]{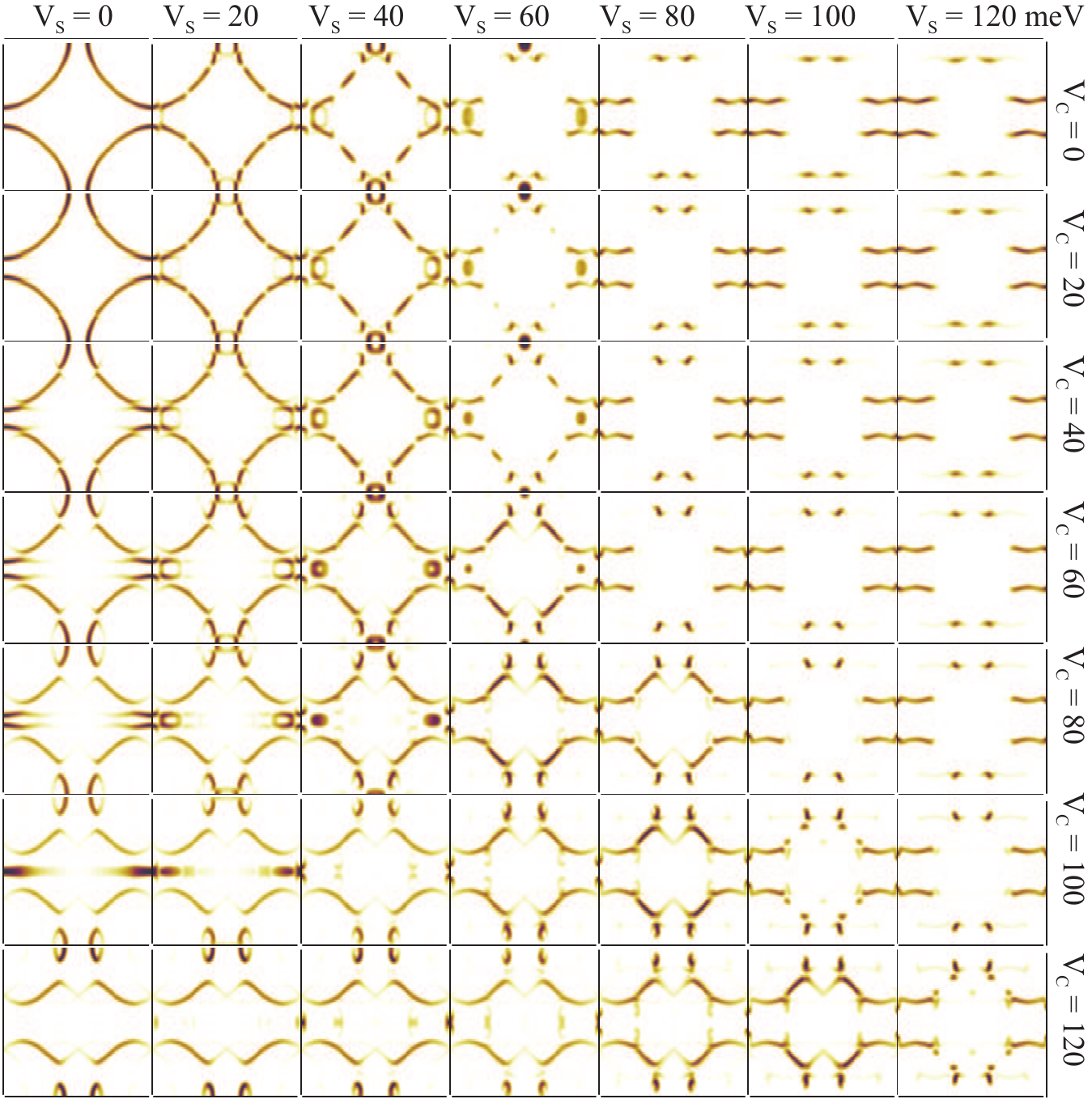}\\
\caption{Calculated FS for various scattering potentials $V_\textup{c}$ and $V_\textup{s}$.
Calculated spectral function was smoothed within small energy and momentum windows
to simulate  experimental resolution.\vspace{-2.0em}}
\end{center}
\end{figure}

From the last formula it can be easily seen that for the case of infinitesimally small potential
$V_{n,m}(\mathbf{q})$ the reconstructed band structure must consist of 8 ``replicas'' obtained from the original
structure shifted by vectors $\mathbf{g}_m$ with the distribution of the spectral weight determined  by the
components of eigenvectors $D_{m, i}(\mathbf{k}-\mathbf{g}_m)$, which results in an infinitesimally small
intensity of all the replicas that do not overlap with the original structure. In case of scattering potential values
comparable to the band width of the original structure the general property of weak replica intensities remains valid,
though the simple ``rule of shifts'' breaks down and the evaluation of the spectral function must be done
numerically.

To check whether the experimental FS can be reproduced within the discussed model, we have
calculated the spectral weight distributions for different spin and charge scattering potentials.
For the unreconstructed band dispersion $\varepsilon_\mathbf{k}  = \varepsilon_0 -2t(\cos k_x +\cos
k_y) - 4 t'\cos k_x \cos k_y  - 2 t'' \cos 2 k_x \cos 2 k_y$ we used experimental data published
for pure LSCO system \cite{Yoshida}, interpolating between doping levels $x=0.07$ and 0.15 and
adjusting the chemical potential to obtain precise doping level of $x=0.125$, which results in the
following parameters: $\varepsilon_0 = 0.175, t=0.25, t'=-0.04,\textup{ and } t'' =
0.02\,\textup{eV} $. The obtained distributions are shown in Fig.\,4. Before comparing  the
calculated FS to the experimental one we have to mention that in the model we have preserved the
anisotropy of the stripe scattering potential, while in the experimental data one is likely to
observe coexistence of stripes running along the $x$ and $y$ direction.  Thus when searching for
the optimal values of $V_\textup{s,c}$ we should not sift out the calculated spectra just because
of their one-dimensionality. From the Fig.\,4 it is obvious that there is no problem for the model
to reproduce the seeming octagonal structure of the experimental FS, though the values of spin
scattering exceeding 80 meV are surely too large. Further restriction on the values of
$V_\textup{S,C}$ can be drawn comparing distances between the most intense  part in the intensity
distributions. Our subjective judgement for the nearest fit to the experimental data would be the
case of $V_\textup{s} \approx 60\pm20$ and $V_\textup{c} \approx 100\pm20$ meV.

     When further comparing the model to the the experimental data the
issue of  the stripe scattering potential periodicity has to be mentioned. This seems to be
especially important for the problem of a pseudogap that  has been observed in  a similar type of
stripe compound \cite{Valla}. Strictly speaking both charge and spin order are not commensurate to
the lattice and have finite correlation lengths \cite{Kim}, which certainly must affect the finer
details in the electronic structure. It has been known for decades from studies of quasicrystals
and quasi-periodic alloys that incommensurate potentials generally lead to suppression of the
spectral weight at the FL, which in that field of research has also been termed as ``pseudogap''
\cite{Wu, Hafner, Ueda}. Moreover there are known successful attempts to transfer the ideas
developed for the quasicrystals onto the problem of mysterious pseudogap in cuprates
\cite{Seibold}. Recently a pseudogap effect, similar to the one discussed in high-$T_\textup{c}$
cuprate, has been detected in the incommensurate phase of CDW bearing compound \cite{Borisenko}
TaSe$_2$ once again suggesting that the incommensurate order might be a long sought common origin
of the pseudogaps. Unfortunately a precise calculation of rational approximates to the
incommensurate structure  is encumbered with drastically increased numeric complexity, although it
is obvious that such a calculation would significantly help to understand the pseudogap formation
mechanism  in high-temperature superconducting cuprates.

This project is part of the Forschergruppe FOR538 and is supported by the DFG under Grants No. KN393/4 and BO1912/2-1.
We thank R. Hübel for technical support.

\end{document}